\documentclass[10pt,twocolumn]{article}
\usepackage[letterpaper,top=0.62in,bottom=0.68in,left=0.64in,right=0.64in,columnsep=0.25in]{geometry}
\usepackage[T1]{fontenc}
\usepackage{newtxtext,newtxmath}
\usepackage{amsmath,bm}
\usepackage{graphicx}
\usepackage{booktabs}
\usepackage{microtype}
\usepackage[numbers,sort&compress]{natbib}
\usepackage[colorlinks=true,citecolor=blue,linkcolor=blue,urlcolor=blue]{hyperref}
\usepackage{doi}
\hypersetup{pdftitle={Full-field and Bloch-periodic-factor discretizations: Accuracy and phantom modes},pdfauthor={Igor Tsukerman}}
\newcommand\figref[1]{Fig.~\ref{#1}}
\newcommand\tableref[1]{Table~\ref{#1}}
\newcommand\sectref[1]{Section~\ref{#1}}
\newcommand{\ii}{\mathrm{i}}
\newcommand{\eps}{\varepsilon}
\newcommand{\bfr}{\mathbf{r}}
\newcommand{\bfq}{\mathbf{q}}

\graphicspath{{figures/}}

\title{Full-field and Bloch-periodic-factor discretizations:\\
Accuracy and phantom modes}

\author{Igor Tsukerman\\[-2pt]
\small Department of Electrical and Computer Engineering, The University of Akron,\\[-2pt]
\small Akron, Ohio 44325-3904, USA\\
\small \href{mailto:igor@uakron.edu}{\texttt{igor@uakron.edu}}}

\date{\small \today}

\begin{document}
\maketitle
\vspace{-1.4em}
\begin{abstract}
One can distinguish four major formulations of the Bloch-mode problem in linear periodic media. First, there are two common complementary ways of slicing the Bloch variety (Bloch wave vectors paired with the corresponding frequencies): frequency vs. wave vector or, conversely, wave vector vs. frequency. This paper deals exclusively with the latter option, directly applicable to lossy media, evanescent waves, and complex band structures. 

A separate crossroads is the full field (FF) vs. the lattice-periodic Bloch factor (PF) formulations. These are fully equivalent on the continuous level, but common discretization techniques may break this equivalence not just approximately but qualitatively. In the FF problem, the primary unknown eigenvalue is the Bloch phase factor (not the wavenumber), which enters only through opposite-boundary coupling. The PF formulation, on the other hand, injects the Bloch wavenumber into the differential operator and leads to a volume quadratic pencil. Consequently, standard PF discretizations of the type considered here, in contrast with the corresponding FF discretizations, violate the reciprocal-lattice (Brillouin-zone-shift) covariance -- which, as the theory and numerical examples in the paper show, may lead to inaccurate or even nonphysical modes.

From the mathematical perspective, the paper highlights the role of structure-preserving algorithms. The practical importance of the results and recommendations stems from the wide adoption of PF formulations in optics, photonics, and beyond -- such as the computation of propagating and evanescent Bloch modes in photonic and acoustic structures, topological photonics, effective-mass and topological insulator band theories.
\end{abstract}
\vspace{-0.6em}

\section{Introduction}\label{sec:Intro}
%
\subsection{Bloch modes -- exact and numerical}\label{sec:Bloch-modes}
Bloch modes, the cornerstone of wave analysis in periodic media, can be written, at a given frequency $\omega$, in phasor form as a lattice-periodic factor times a complex exponential with a Bloch wave vector $\bfq$. In general, the frequency -- wave-vector pairs $(\omega, \bfq)$ form a multidimensional complex Bloch variety \cite{EngstromRichter2009,Engstrom2014,LacknerMengMonk2019}.

In practice, this variety can be sampled as $\omega(\bfq)$, with $\bfq$ in the first Brillouin zone (FBZ) \cite{Joannopoulos2008}, or as $\bfq(\omega)$. The $\omega$-vs.-$\bfq$ formulation is Hermitian for simple lossless media, but becomes nonlinear for frequency-dispersive media and, moreover, is not naturally adapted to evanescent modes in band gaps, for which $\bfq$ is complex. For rational Drude--Lorentz dispersion, the nonlinear frequency dependence can instead be eliminated by augmenting Maxwell's equations with auxiliary polarization variables. This auxiliary-field construction has a long history in time-domain treatments of dispersive media \cite{KashiwaFukai1990,JosephHagnessTaflove1991} and was subsequently developed in Hamiltonian formulations of dispersive and absorptive electrodynamics \cite{Tip1998,BhatSipe2006}. Using these ideas, the photonic-band problem for lossless dispersive periodic media can be cast in Hermitian form \cite{RamanFan2010}.

\textit{The present paper deals exclusively with the $\bfq(\omega)$ problem}, which is non-Hermitian and directly applicable to lossy media, leaky modes, and evanescent waves \cite{EngstromRichter2009,Engstrom2014,LacknerMengMonk2019,HuangLiLinLinTian2020,NotarosPopovic2015}. This problem admits two natural formulations (\sectref{sec:1D-toy-formulations}): in terms of either the full field (FF) subject to Bloch boundary conditions or, alternatively, the lattice-periodic factor (PF). The latter is arguably more common. The periodic factors themselves are central in topological photonics \cite{Vanderbilt2018,Tsukerman2025} and condensed-matter theories \cite{LuttingerKohn1955,Kane1957,FuKane2007}.

At the continuous level, the two formulations are equivalent. Consider the 1D case to start with. If the lattice period is $a$ and the reciprocal-lattice constant is $\kappa = 2 \pi/a$, values of $q$ that differ by an integer multiple of $\kappa$ are aliases corresponding to exactly the same Bloch mode. We shall refer to this as exact reciprocal (or Brillouin-zone) covariance.

However, common fixed-grid discretizations, such as the finite-difference (FD) and finite-element (FE) PF schemes considered here, can break this covariance; \textit{it is one of the central issues of the paper}. Indeed, these PF schemes inject the wavenumber $q$ (or, more generally, the wave vector) into the differential operator. As a result, discretization produces a volume polynomial pencil (Sections~\ref{sec:1D-toy-matrix} and \ref{sec:2D-formulation}). FF schemes are quite different in that regard: the primary unknown eigenvalue is not the wavenumber but the Bloch phase multiplier $z = \exp(\ii q a)$. This multiplier enters only through the boundary coupling, while the interior operator is independent of $q$. Thus, the standard FF discretizations considered here preserve reciprocal covariance by construction, whereas their PF counterparts do not. A specially designed structure-preserving PF discretization is not excluded by this argument. Importantly, this is not a purely academic matter; as the paper shows, the computed PF eigenmodes can become highly inaccurate or even nonphysical, even though the underlying numerical schemes are well-established and solid.

The remainder of the paper is organized as follows.
\sectref{sec:1D-toy} introduces a 1D toy problem, for which the appearance and causes of the phantom numerical modes can be investigated fully. \sectref{sec:1D-SiC} enhances the 1D model with a lossy dispersive layer in the lattice cell; an exact solution is still available for analysis and verification. \sectref{sec:2D-problem} examines a 2D periodic structure with dispersive lossy bars in the lattice cell. The findings, conclusions and recommendations appear in \sectref{sec:discussion}.
%
\subsection{Existing analyses and simulations}\label{sec:Literature-review}
%
The following survey, while short and not exhaustive, helps to put this paper's analysis in proper context.

A scalar PF quadratic eigenproblem discretized by finite elements is examined in \cite{EngstromRichter2009}. The computations target selected eigenvalues near prescribed Krylov shifts; additional physical filters are applied. Convergence of isolated eigenvalues under appropriate hypotheses is established in the follow-up study \cite{Engstrom2014}. The present paper is not in conflict with \cite{EngstromRichter2009,Engstrom2014}; its focus is on the pollution of the spectrum with inaccurate or nonphysical modes and the principal difference between the discretized FF and PF formulations.

A PF finite-element $q(\omega)$ formulation for complex bands of a dispersive plasmonic crystal is presented in \cite{DavancoUrzhumovShvets2007}. The authors report duplicate out-of-FBZ eigensolutions whose $\operatorname{Re}q$ differs from that of an FBZ root by a reciprocal lattice vector while $\operatorname{Im}q$ is identical. Thus their published second-order FEM spectra resolve low-order reciprocal copies well enough to be recognizable as aliases. The published results do not establish whether higher-order reciprocal copies were computed or retained. Part of the same complex-band diagram was subsequently reproduced with the Flexible Local Approximation Method (FLAME) \cite{Tsukerman2009QuasiHomogeneous}, which by construction utilizes the FF formulation.

A 3D staggered-grid simulation \cite{HuangLiLinLinTian2020} is also selective with respect to the eigensolutions it yields. Namely, its shift-and-invert iteration returns roots near prescribed shifts or near the unit circle, rather than the entire raw spectrum. 

A $q(\omega)$ finite-difference formulation for selected radiative waveguide and grating modes is presented in \cite{NotarosPopovic2015}. These are open systems whose legitimate leaky and evanescent states already populate the complex plane and are not in a one-to-one correspondence with the spectra considered in this paper.

The phantom modes discussed here are distinct from the classical spurious modes of vector finite-element electromagnetics. In the latter case, the mechanism is associated with violation of the zero-divergence condition in nodal element approximations \cite{Bossavit1998}, while rigorous mathematical theories revolve around discrete compactness \cite{Monk2003,BoffiDemkowiczCostabel2003}. Effects considered in the present paper are much more down-to-earth.
%
\section{A toy problem: 1D empty cell}\label{sec:1D-toy}
%
\subsection{Continuous formulations}\label{sec:1D-toy-formulations}
%
As noted in the introduction, the breaking of reciprocal covariance and the appearance of phantom modes can be exposed already in the toy model of a 1D homogeneous lattice cell. 

Gaussian units and the phasor convention $\exp(-\ii\omega t)$ are used throughout. Consider the Bloch problem in a homogeneous cell $0 \leq x \leq a$:
\begin{equation}\label{eqn:1D-full}
   E''(x) \,+\, k^2 E(x) ~=~ 0,
   \quad 
   k = \text{const} \,>\, 0
\end{equation}
\begin{equation}\label{eqn:1D-Bloch-conditions}
   E(a) \,=\, z E(0),
   \quad
   E'(a) \,= z E'(0),
   \quad
   z \,=\, \exp(\ii qa)
\end{equation}
In this trivial case, the phase multipliers \( z \) are
\begin{equation}\label{eqn:1D-full-field-Bloch-roots}
   z_\pm = \exp(\pm \ii k a),
   \quad
   q_\pm = \pm k ~~ (\bmod \ \kappa).
\end{equation}
Writing
\begin{equation}\label{eqn:E-eq-Etilde-exp-iqx}
   E(x) ~=~ \widetilde E(x) \exp(\ii q x)
\end{equation}
gives
\begin{equation}\label{eqn:1D-periodic-continuous}
   L(q)\widetilde E \,:=\,
   \widetilde E'' \,+\, 2 \ii q \widetilde E'
   \,+\, ( k^2 - q^2 ) \widetilde E ~=~ 0,
   \quad
   \widetilde E(x + a) \,=\, \widetilde E(x).
\end{equation}
The set of aliased unit-amplitude solutions is
\begin{equation}\label{eqn:1D-exact-aliases}
   \widetilde E_m(x) = \exp(\ii m \kappa x),
   \quad
    q_{m,\pm} =\, \pm k - m \kappa,
   \quad
   m \in \mathbb Z
\end{equation}
All integers $m$ produce the same two full fields. Formally, 
\begin{equation}\label{eqn:continuous-reciprocal-covariance}
   L(q + m \kappa) ~=~ T_m^{-1}L(q)T_m \quad
   \forall m \in \mathbb Z
\end{equation}
where
\begin{equation}\label{eqn:Tm-exp-imkappa}
   T_m f \,=\, \exp(\ii m \kappa x) f
\end{equation}
for $a$-periodic functions f.

This reciprocal covariance is a feature worth preserving in numerical models and is one of the central issues of the paper.
%
\subsection{Discretization}\label{sec:1D-toy-discretization}
%
Discretization may alter the aliasing structure qualitatively. As will become clear in the remainder, the consequences can be significant for many discretization techniques, such as standard FD schemes or classical FEM. 

For concreteness and simplicity, consider central-difference schemes. Divide the cell into $N$ intervals, $h = a/N$. There are $N + 2$ grid nodes $x_j = (j - 1) h$, $j = 1,\ldots,N + 2$: $E_1,\ldots,E_N$ are master values, while $E_{N + 1}$ and $E_{N + 2}$ are slave values introduced to impose the Bloch boundary conditions. The scheme for the FF equation \eqref{eqn:1D-full} is
\begin{equation}\label{eqn:1D-full-field-FD}
   \frac{E_{j + 1} - 2 E_j + E_{j - 1}}{h^2} \,+\, k^2 E_j \,=~ 0,
   \qquad j = 2, \ldots, N + 1,
\end{equation}
with the boundary conditions
\begin{equation}\label{eqn:1D-full-field-FD-bc}
   E_{N + 1} = z E_1,
   \quad
   E_{N + 2} = z E_2,
   \quad
   z \,=\, \exp(\ii q a)
\end{equation}
where $z$ and $q$ are the Bloch multiplier and wavenumber, respectively.

The standard characteristic equation for the central-difference recurrence \cite{Samarskii2001,Collatz1960} corresponding to \eqref{eqn:1D-full-field-FD} is
\begin{equation}\label{eqn:1D-rho-characteristic}
   \rho^2 - ( 2 - k^2 h^2 ) \rho + 1 = 0,
   \quad
   \text{roots} ~ \rho_{1,2}
\end{equation}
Equivalently, setting $\rho=\exp(\ii \xi h)$ gives the familiar centered-second-difference Fourier symbol $-4h^{-2}\sin^2(\xi h/2)$.

This leads to exactly one numerical multiplier pair
\begin{equation}\label{eqn:1D-zh}
   z_{h1,2} ~=~ \rho_{1,2}^N  ~:=~ \exp(\ii q_{h \pm}a)
\end{equation}
For $k h < 2$, direct algebra gives
\begin{equation}\label{eqn:1D-full-error}
   q_{h,\pm} \,=\, \pm \left( k \,+\, \frac{k^3 h^2}{24} 
   \,+\, \mathcal{O} \bigl( h^4 \bigr) \right).
\end{equation}
Now, let us contrast this with the central discretization of the PF problem \eqref{eqn:1D-periodic-continuous}:
\begin{multline}\label{eqn:1D-periodic-FD}
   \frac{\widetilde E_{j + 1} - 2 \widetilde E_j + \widetilde E_{j - 1}}{h^2}
   \,+\, \frac{\ii q}{h} ( \widetilde E_{j + 1} - \widetilde E_{j - 1} ) \\
   \,+\, ( k^2 - q^2 ) \widetilde E_j = 0,
   \qquad
   \widetilde E_{\alpha + N} = \widetilde E_\alpha, ~~ \alpha = 1,2.
\end{multline}
For an integer reciprocal shift $m\kappa$, define sampled multiplication by
$(T_{m,h}v)_j=\exp(\ii m\kappa x_j)v_j$ and the central first-difference operator by
$(D_hv)_j=(v_{j+1}-v_{j-1})/(2h)$. The latter does not satisfy the exact gauge-product identity
$D_hT_{m,h}=T_{m,h}(D_h+\ii m\kappa I)$. Consequently the exact similarity relation does not generally hold on the discrete level:
\begin{equation}\label{eqn:discrete-reciprocal-covariance-defect}
   Q_h(q+m\kappa) ~\neq~ T_{m,h}^{-1} \, Q_h(q) \, T_{m,h}.
\end{equation}
This lack of reciprocal covariance may be viewed as the root cause of the qualitative difference between the FF and PF schemes \eqref{eqn:1D-full-field-FD} and \eqref{eqn:1D-periodic-FD}, respectively. 

More specifically, problem \eqref{eqn:1D-periodic-FD} has \textit{multiple} solutions, easily found via discrete Fourier transform. For \textit{each} grid harmonic $\widetilde E_j = \exp(2 \pi \ii m j/N)$, the two dimensionless roots $\widetilde q = q a$ are
\begin{equation}\label{eqn:1D-qm}
   \widetilde q_{m, \pm}^{(N)} ~=~ - N \sin \widetilde m
   \, \pm \, \sqrt{\widetilde k^{\,2} - 4 N^2 \sin^4 ( \widetilde m/2 )},
\end{equation}
where $\widetilde k = k a$ and $\widetilde m = 2 \pi m/N$. For $ka \leq \pi$, the $m = 0$ harmonic gives the physical FBZ pair exactly. Every nonzero grid harmonic contributes another pair.

Scheme \eqref{eqn:1D-periodic-FD} is second-order consistent, but not uniformly so in $m$. Evaluate Eq.~\eqref{eqn:1D-periodic-FD} on the sampled exact alias pair
$\widetilde E_j = \exp(2 \pi \ii m j/N)$ and
$q_{m,\pm} = \pm k - m \kappa$. Dividing the resulting nodal residual by $\widetilde E_j$ and normalizing gives
\begin{equation}\label{eqn:alias-consistency}
   \widehat\tau_{m,\pm} ~=~ \frac{1}{N^2}
   \left[ \pm \frac{\widehat k ( 2 \pi m )^3}{3}
   - \frac{( 2 \pi m )^4}{4} \right] + \text{h.o.t.},
\end{equation}
where $\widehat k = k a$, and ``h.o.t.'' stands for higher-order terms. The terms in the brackets grow as $m^3$ and $m^4$, while the denominator grows only as $N^2$. As the FD grid is refined, the roots \eqref{eqn:1D-qm} corresponding to each fixed $m$ converge, but at the same time new high-order harmonics emerge and produce inaccurate roots.
%
\subsection{Matrix structure and Schur reduction}\label{sec:1D-toy-matrix}
%
The Fourier algebra above is simple for an empty cell; to extend the analysis to heterogeneous cells and higher dimensions, it is useful to recast Eq.~\eqref{eqn:1D-full-field-FD} and the Bloch relations into matrix form. With the master and slave values defined above, let $\underline E$ collect all $N + 2$ nodal values. The resulting equations can be written as a generalized eigenproblem 
\begin{equation}\label{eqn:LE-eq--zZE}
   L \underline E ~=~ z \, Z \underline E
\end{equation}
Schur elimination \cite{HornJohnson2013} of the $N$ master values reduces this problem of order $N + 2$ to just a $2 \times 2$ pencil. This produces exactly two roots, in accord with the characteristic equation \eqref{eqn:1D-rho-characteristic}.

The PF discretization, on the other hand, has the quadratic form
\begin{equation}\label{eqn:QEP}
   Q(q) \, \underline{\widetilde E} ~:=~
   \left( Q_\Delta + k_0^2 Q_\eps + 2 \ii q Q_x + q^2 Q_2 \right)
   \underline{\widetilde E}
   ~=~ 0.
\end{equation}
Here $Q_\Delta$, $Q_\eps$, $Q_x$, and $Q_2$ represent the four respective terms in \eqref{eqn:1D-periodic-FD}. In contrast with the FF system, all matrices correspond to the interior of the cell. The algebraic order of this problem grows as the grid is refined, leading to multiple additional roots. For each fixed reciprocal harmonic these roots approach the corresponding continuous aliases, but this convergence is not uniform with respect to the harmonic order $m$, as the asymptotic relation \eqref{eqn:alias-consistency} indicates. Inaccurate or nonphysical roots keep polluting the spectrum in the process of grid refinement.

This can be illustrated with a simple numerical example. Let $a=1$, $N=40$, and $a/\lambda=0.1$, so $ka=0.2\pi$. The only physical roots are $qa \approx \pm 0.628319$ for $m=0$. For $m=1$, Eq.~\eqref{eqn:1D-qm} gives
\begin{equation}\label{eqn:1D-selected-phantom-root}
   qa \approx -5.867172 \approx 0.416013
   ~~
   (\bmod{2\pi}).
\end{equation}
which is not, at a reasonable level of precision, an alias of $\pm 0.628319$.

\begin{figure}
\centering
\includegraphics[width=0.92\columnwidth]{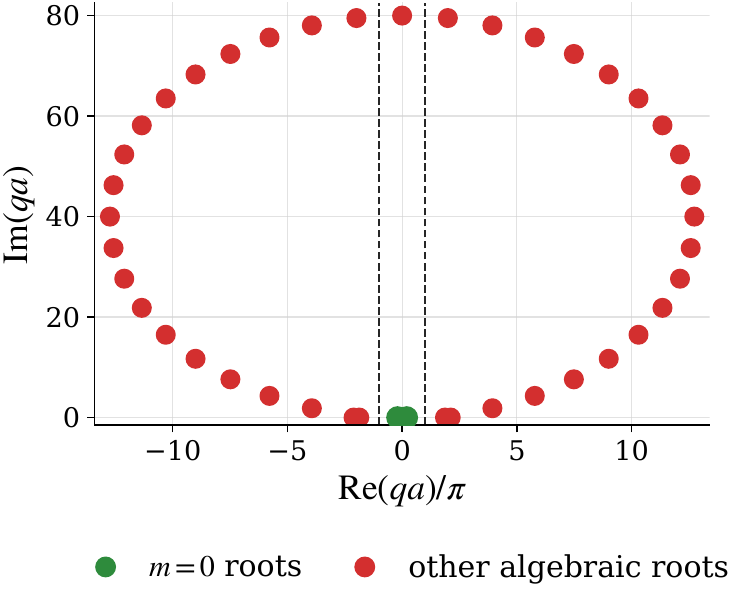}
\includegraphics[width=0.97\columnwidth]{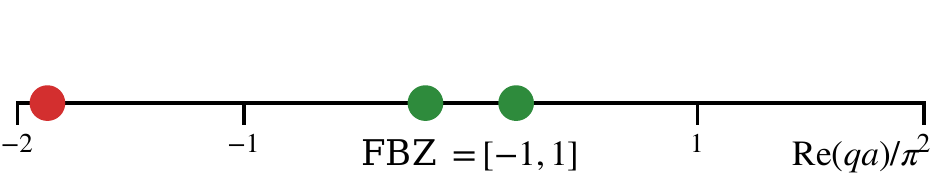}
\caption{Unfiltered PF spectrum for the one-dimensional empty cell, with $N=40$ and $ka=0.2\pi$. Top: the non-growing half-plane, $\operatorname{Im}(qa)\geq 0$, of the finite algebraic spectrum of the discrete problem \eqref{eqn:1D-periodic-FD}. Green points are the physical $m=0$ roots and red points are the nonphysical ones. Bottom: the two physical real roots (green) and the phantom $m=1$ root (red).}
\label{fig:empty-cell-spectrum}
\end{figure}

\begin{figure}
\centering
\includegraphics[width=0.98\columnwidth]{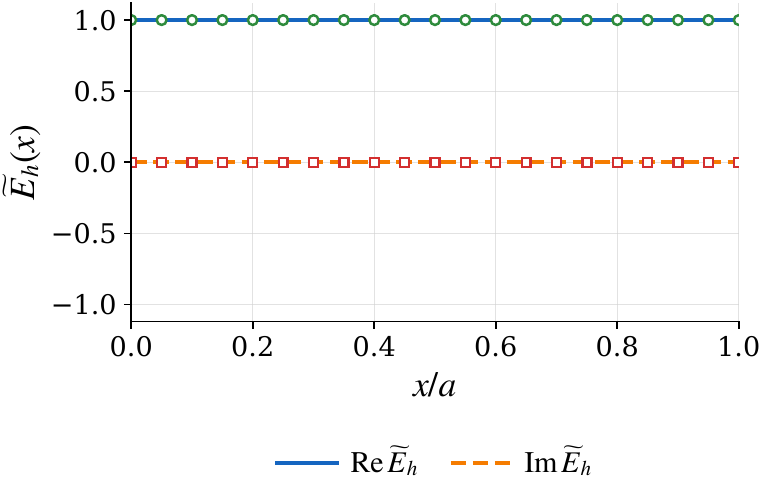}\par
\includegraphics[width=0.98\columnwidth]{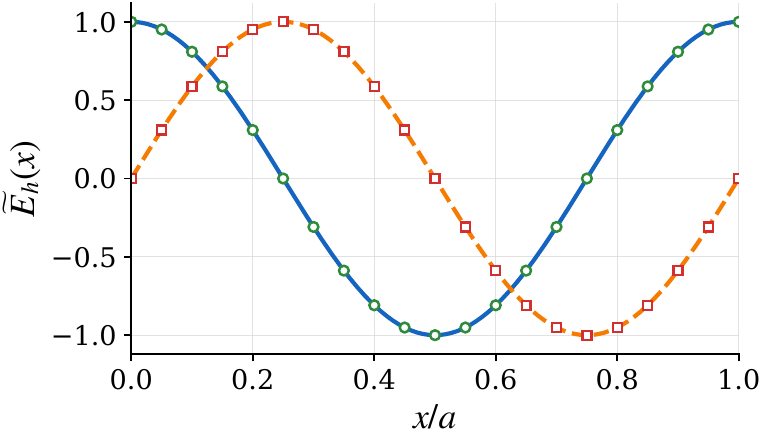}
\caption{One-dimensional periodic factors $\widetilde E_h$ for $a=1$, $N=40$, and $a/\lambda=0.1$. Markers show grid values and the curves are guides to the eye. The physical (top) and nonphysical (bottom) factors corresponding to $m=0$ and $m=1$, respectively.}
\label{fig:1D-physical-nonphysical}
\end{figure}

The phantom nature of the root \eqref{eqn:1D-selected-phantom-root} is further illustrated with Figs.~\ref{fig:empty-cell-spectrum} and \ref{fig:1D-physical-nonphysical}. The top panel of \figref{fig:empty-cell-spectrum} shows the non-growing half-plane of the raw spectrum of the discrete PF problem \eqref{eqn:1D-periodic-FD}. Each grid harmonic $m$ produces a pair of roots. The non-real roots cannot represent physical evanescent modes, which are in fact absent in a homogeneous lossless cell. The bottom panel shows the physical real roots and the phantom $m=1$ root.

\figref{fig:1D-physical-nonphysical} compares the periodic factors themselves. For $m=0$, this factor is (correctly) constant across the cell; but the $m=1$ exhibits a nonphysical yet grid-resolved oscillation.

%
\section{Example: a lossy dispersive layer}\label{sec:1D-SiC}
%
\subsection{1D model}
As we have seen, the appearance of nonphysical modes is already visible in the toy empty-cell model. To explore the related effects further, let us turn to wave propagation in a layered lossy dispersive medium, still keeping the model one-dimensional.

For illustration, consider a SiC layer in an otherwise empty 1D lattice cell $[0, a]$. For purposes of this paper, the choice of SiC is discretionary and  loosely motivated by (i) its wide range of losses, from quite low to high, which is instructive for numerical studies, and (ii) the availability of an accurate analytical representation of the permittivity as a function of frequency \eqref{eqn:SiC-epsilon}. It may be noted in passing -- beyond the scope of this paper -- that SiC is standard in mid-infrared phonon-polariton photonics. Its Reststrahlen band, the phonon-resonance interval between the transverse-optical (TO) and longitudinal-optical (LO) frequencies, has been exploited for strong field enhancement and for critically coupled surface-phonon-polariton excitation, with applications in compact infrared components \cite{HillenbrandTaubnerKeilmann2002,NeunerEtAl2009,BrehmSchliesserCajkoTsukermanKeilmann2008}. The standard analytical solution via the transfer-matrix method for waves in linear layered structures \cite{Yeh2005} is available for verification of the numerical results. For purposes of this paper, it is sufficient to consider propagation perpendicular to the layers; no auxiliary transverse phase constant is introduced. 

In the numerical experiments, the cell size is set to $a = 2.5 \, \mu$m, representative of the mid-infrared scale; the width of a centered SiC layer is $3a/8$. 

The spectroscopic wavenumber is conventionally defined as $\nu = \omega/(2 \pi c) = \lambda^{-1}$ and expressed in $\mathrm{cm}^{-1}$. The Lorentz model for the permittivity is
\begin{equation}\label{eqn:SiC-epsilon}
   \eps_{\mathrm{SiC}}(\nu) ~=~ \eps_\infty \,
   \frac{\nu_{\mathrm{LO}}^2 - \nu^2 - \ii \gamma \nu}
   {\nu_{\mathrm{TO}}^2 - \nu^2 - \ii \gamma \nu},
\end{equation}
where, up to the quoted significant digits, $\eps_\infty = 6.7$, $\nu_{\mathrm{TO}} = 793\,\mathrm{cm}^{-1}$,
$\nu_{\mathrm{LO}} = 969\,\mathrm{cm}^{-1}$, and
$\gamma = 4.76\,\mathrm{cm}^{-1}$ \cite{SpitzerKleinmanWalsh1959}; the subscripts TO and LO denote the transverse-optical and longitudinal-optical phonon frequencies, respectively.

Both FF and PF control-volume schemes use the same material-fitted grids. Because the SiC interfaces lie at $x/a=5/16$ and $11/16$, taking $N$ to be a multiple of 16 places both interfaces on grid nodes. This reduces the numerical errors inherent in the schemes but unrelated to the subject of this paper. The PF problem leads to the quadratic pencil \eqref{eqn:QEP}. 

\begin{figure}
\centering
\includegraphics[width=0.98\columnwidth]{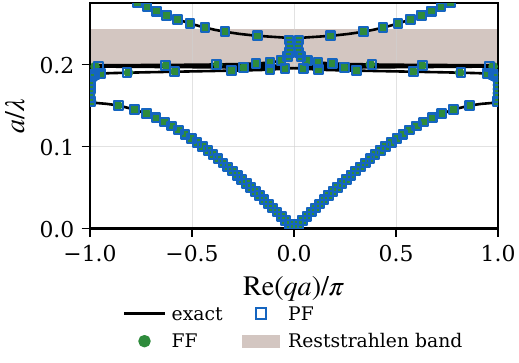}
\includegraphics[width=0.98\columnwidth]{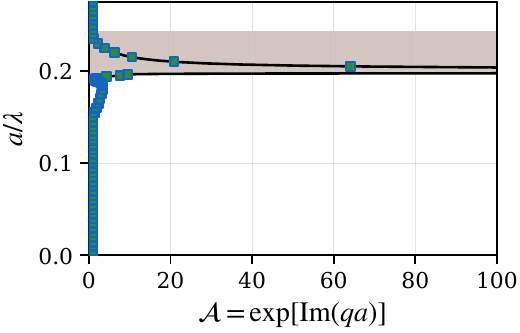}
\caption{Complex bands of the 1D lossy-SiC structure over $0 \leq a/\lambda \leq 0.2750$; $N=32$. Top: normalized frequency $a/\lambda$ versus $\operatorname{Re}(qa)/\pi$. Bottom: $a/\lambda$ versus the amplitude attenuation factor per period $\mathcal A=\exp[\operatorname{Im}(qa)]$, with the horizontal axis limited to $0 \leq \mathcal A \leq 100$. Results are marked as follows. Exact transfer-matrix -- black curves, FF -- green filled circles, PF -- blue open squares. The light-brown horizontal interval is the SiC Reststrahlen band. The central legend applies to both panels.}
\label{fig:SiC-1D-complex-bands}
\end{figure}

The complex band diagram, shown in \figref{fig:SiC-1D-complex-bands} for reference, covers $0$--$1100\,\mathrm{cm}^{-1}$, or $0 \leq a/\lambda \leq 0.2750$. The physical branches are shown continuously, while markers are subsampled for readability. The attenuation outside the Reststrahlen band is small but not zero.

The top panel displays the opposite directions symmetrically. The bottom panel uses the amplitude attenuation factor per period, $\mathcal A=\exp[\operatorname{Im}(qa)]$, on the decaying branch and is shown up to $\mathcal A=100$. The analytical and numerical results agree closely over this practical range.
The nearly horizontal exact segment just below the lower Reststrahlen edge is not a flat band. It is the steep lower-frequency flank of the TO resonance: between $789$ and $790\,\mathrm{cm}^{-1}$, the transfer-matrix result changes from $\mathcal A\approx48$ to $\mathcal A\approx113$, while $a/\lambda$ changes only from $0.19725$ to $0.19750$. 

The central issue is the qualitative distinction of physical and nonphysical modes, as illustrated by, e.g., \figref{fig:1D-physical-nonphysical}; regular errors of FD schemes are well-understood and of lesser interest from the physics perspective. Nevertheless, it is worth mentioning that these errors depend on the level of attenuation and affect the FF and PF results differently. \tableref{tab:SiC-1D-accuracy} lists three representative cases for $N = 64$. At $755\,\mathrm{cm}^{-1}$, the PF attenuation error is about three times smaller than that of the FF. Near the transverse-optical resonance, $795\,\mathrm{cm}^{-1}$, the FF results are better. But for the frequency just $2\,\mathrm{cm}^{-1}$ higher ($797\,\mathrm{cm}^{-1}$), the PF solution is again substantially more accurate. Refinement computations (not shown) confirm approximately second-order convergence for both FF and PF formulations. These observations suggest that factoring out the Bloch exponential can help approximate strongly evanescent fields but can also be disadvantageous in a narrow resonant regime. 

One additional observation is relevant to reciprocal covariance. At $795\,\mathrm{cm}^{-1}$ and $N=64$, the direct-FBZ PF root is $qa \approx -1.833267+10.722377\ii$, with complex-$q$ error $5.60 \times 10^{-2}$. An exterior PF root, $qa=4.510556+10.679256\ii$, is congruent modulo $2\pi$ to $-1.772629+10.679256\ii$ and happens to have the smaller error $1.85\times10^{-2}$. Reciprocal relabeling is not a remedy for a nonphysical root: when the periodic factor is transformed consistently, the full Bloch field is unchanged. This example instead shows that reciprocal representatives can have substantially different accuracy in discrete PF problems.

\begin{table}
\centering
\caption{Selected one-dimensional SiC attenuation errors at $N = 64$. Here $\alpha = |\operatorname{Im}(qa)|$.}
\label{tab:SiC-1D-accuracy}
\begin{tabular}{c|c|c|c}
\toprule
$\nu$ (cm$^{-1}$) & $\alpha$ & FF & PF\\
\midrule
755 & 0.775 & $4.83\times10^{-3}$ & $1.74\times10^{-3}$\\
795 & 10.691 & $4.68\times10^{-3}$ & $3.17\times10^{-2}$\\
797 & 9.364 & $1.31\times10^{-2}$ & $2.66\times10^{-4}$\\
\bottomrule
\end{tabular}
\end{table}
\vspace{0.6em}

\section{A 2D example}\label{sec:2D-problem}
%
\subsection{Formulation}\label{sec:2D-formulation}
%
The 1D model exposes the lack of reciprocal covariance in the discrete PF problem, with related numerical errors and phantom modes. The 2D case illustrates the same effects in a larger dispersive problem.

The governing equation for the electromagnetic $s$ mode in a square cell of side $a$ is, in the frequency domain,
\begin{equation}\label{eqn:2D-Helmholtz}
   \nabla^2 E \,+\, k_0^2 \eps(\bfr) E ~=~ 0
\end{equation}
For concreteness, consider Bloch waves propagating in $x$, with periodicity in $y$:
\begin{equation}\label{eqn:2D-Bloch-decomposition}
   E(\bfr) = \widetilde E(\bfr) \exp(\ii q_x x).
\end{equation}
The periodic factor satisfies
\begin{equation}\label{eqn:2D-periodic-factor}
   \nabla^2 \widetilde E \,+\, 2 \ii q_x \partial_x \widetilde E
   \,+\, \left( k_0^2 \eps - q_x^2 \right) \widetilde E ~=~ 0.
\end{equation}
Flux-balance / control-volume discretization \cite{Samarskii2001,Collatz1960,Tsukerman2026} gives a volume quadratic pencil, again of the generic form \eqref{eqn:QEP}.

For the FF formulation, let $\underline E_b$ denote nodal values on one vertical side and $\underline E_i$ all remaining values. After the values on the opposite side have been related by the multiplier $z$, the discrete equations assume the form
\begin{equation}\label{eqn:2D-full-field-block-system}
   \begin{aligned}
      A_{ii} \underline E_i \,+\, A_{ib}(z) \underline E_b &~=~ 0, \\
      A_{bi}(z) \underline E_i \,+\, A_{bb}(z) \underline E_b &~=~ 0.
   \end{aligned}
\end{equation}
Eliminating the interior values gives the boundary-sized quadratic pencil
\begin{equation}\label{eqn:2D-Schur}
   ( S_{-1} + S_0 z + S_1 z^2 ) \underline E_b = 0.
\end{equation}
For finite $z$ and nonsingular interior block $A_{ii}$, this Schur reduction transforms the problem to an equivalent one. This may be convenient for the analysis and computation, but does not affect the modes. The critical difference with \eqref{eqn:2D-periodic-factor} lies not in the Schur elimination but in the fact that $z$, and not $q_x$, enters the FF equation.

Plane-wave expansion (PWE) is standard and used below for  comparison. The $\omega(q)$ and the complex $q(\omega)$ PWE problems are described, for example, in \cite{Sakoda2005} and \cite{ChoUshidaBamba2005,HsueFreemanGu2005}, respectively, among many other places. In the $q(\omega)$ case, one again ends up with a quadratic pencil:
\begin{equation}\label{eqn:PWE-q-omega}
   \left[ q_x^2 I \,+\, 2 q_x G_x \,+\, G_x^2 \,+\,
   G_y^2 \,-\, k_0^2 \mathcal{E} \right]
   \underline{\widetilde E} = 0.
\end{equation}
Here $G_x$ and $G_y$ are diagonal matrices of the reciprocal-vector components, and $\mathcal{E}$ is the Fourier-convolution matrix of the permittivity. PWE provides a useful independent approximation, which may or may not be better than that of FD schemes, depending on various physical, geometric and numerical parameters. Generally speaking, finite differences and finite elements retain their usual advantages for representing local geometric features and interfaces without the Gibbs oscillations.
%
\subsection{Geometry and numerical setup}\label{sec:2D-setup}
\begin{figure}
\centering
\includegraphics[width=0.90\columnwidth]{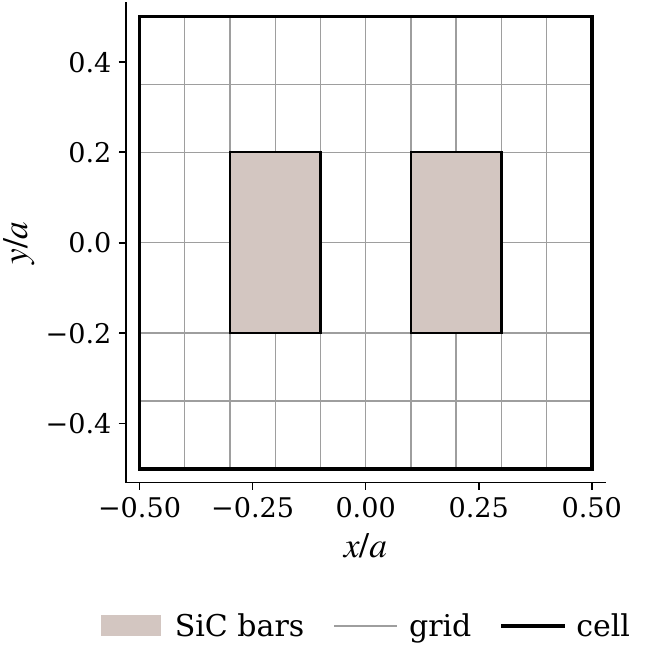}
\caption{Two-SiC-bar unit cell and the material-fitted $6 \times 10$ grid.}
\label{fig:two-bar-geometry}
\end{figure}

%
The square cell is $[-a/2,a/2]^2$, with $a = 2.5\,\mu\mathrm{m}$. Two SiC bars occupy the rectangles $[-0.3a, -0.1a] \times [-0.2a, 0.2a]$ and
$[0.1a, 0.3a] \times [-0.2a, 0.2a]$; the host is air. The SiC permittivity is given by Eq.~\eqref{eqn:SiC-epsilon}. Figure~\ref{fig:two-bar-geometry} shows the material-fitted $6 \times 10$ grid; uniform subdivision by factors of two and three gives the $12 \times 20$ and $18 \times 30$ grids used for refinement checks.

The frequency sweep is $650$--$1100\,\mathrm{cm}^{-1}$, corresponding to $0.1625 \leq a/\lambda \leq 0.2750$, with denser sampling near the transverse-optical resonance. The FF calculation uses the Schur pencil \eqref{eqn:2D-Schur}, whereas the PF calculation uses the discretized volume pencil \eqref{eqn:QEP}. FF multipliers $z$ are reported through their principal logarithmic labels in the FBZ.

\begin{figure}
\centering
\includegraphics[width=0.98\columnwidth]{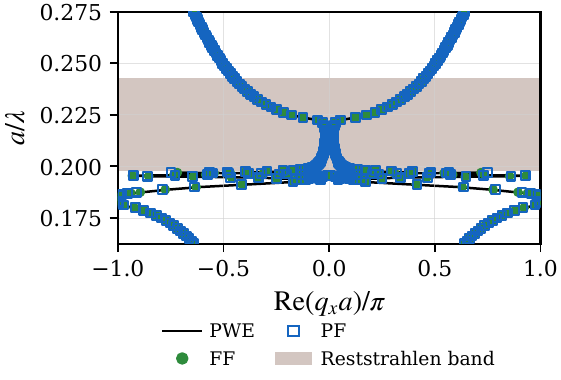}\par
\includegraphics[width=0.98\columnwidth]{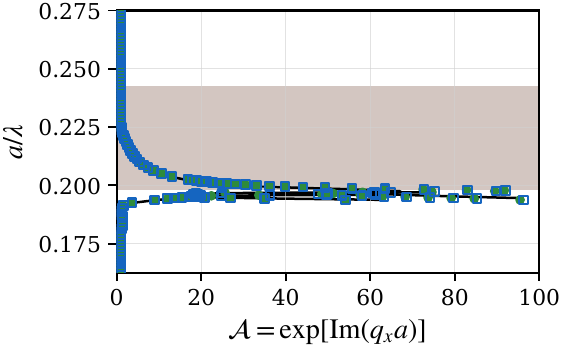}
\caption{Reference complex-band diagram of the two-SiC-bar crystal. Top: $\operatorname{Re}(q_xa)/\pi$, confined to the FBZ. Bottom: amplitude attenuation factor per period $\mathcal A=\exp[\operatorname{Im}(q_xa)]$, with the horizontal axis limited to $0\leq\mathcal A\leq100$. Black curves are PWE, green filled circles are FF, and blue open squares are PF. The shaded interval is the SiC Reststrahlen band. The single legend beneath the top panel applies to both panels.}
\label{fig:two-bar-raw}
\end{figure}

Figure~\ref{fig:two-bar-raw} provides a reference complex-band diagram. The phase panel is confined to the FBZ, while the attenuation panel uses the amplitude attenuation factor per period, $\mathcal A=\exp[\operatorname{Im}(q_xa)]$, and is limited to $\mathcal A\leq100$. The PWE, FF, and PF calculations show the same band structure and agree closely over most of the displayed range; the largest visible differences occur in the rapid variation near the transverse-optical resonance.

The apparent swarm near the lower Reststrahlen edge has the same origin as the nearly horizontal feature in the one-dimensional example. In the narrow neighborhood of the TO resonance, the Lorentz permittivity varies very rapidly with frequency and the phase constants sweep through a substantial fraction of the FBZ while $a/\lambda$ changes only slightly. 

\subsection{Physical and nonphysical roots under refinement}\label{sec:2D-splitting}
%
A representative coarse-grid anomaly occurs at $650\,\mathrm{cm}^{-1}$. On the $20\times12$ grid the PF pencil produces the exterior root
\[
q_xa=7.93026+0.00921\ii .
\]
Its FBZ alias is $1.64707+0.00921\ii$, whereas the PF root computed directly there is $2.01315+0.00824\ii$. Reciprocal relabeling does not repair this discrepancy: when the periodic factor is transformed consistently, the full Bloch field is unchanged, so the exterior root remains the same coarse-grid numerical mode. This is analogous to the $m=0$ and $m=1$ periodic-factor solutions in \figref{fig:1D-physical-nonphysical}.

\begin{figure}
\centering
\includegraphics[width=0.98\columnwidth]{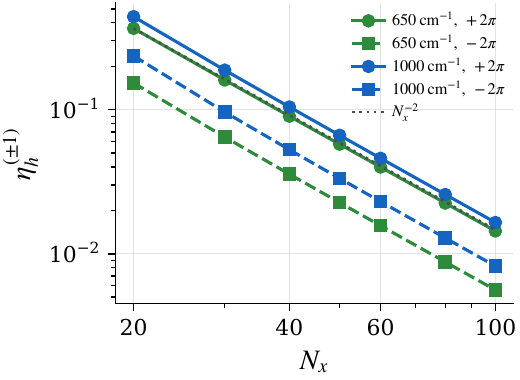}
\caption{Differences between the computed FBZ PF roots and their out-of-FBZ $\pm 2\pi$ aliases at wavenumbers $650$ and $1000\,\mathrm{cm}^{-1}$. The reciprocal-covariance defect decreases approximately at second order under refinement.}
\label{fig:two-bar-alias-refinement}
\end{figure}

Under refinement, this particular anomaly does not persist as a separate mode. Figure~\ref{fig:two-bar-alias-refinement} shows the exterior $\pm2\pi$ PF representatives approaching the directly computed first-zone PF root at both $650$ and $1000\,\mathrm{cm}^{-1}$, with an approximately second-order decrease of the splitting. Thus this coarse-grid phantom tends to a physical reciprocal representative. However, other coarse-grid phantom roots continue to appear outside the FBZ as the volume PF spectrum expands with the grid; they add no new physical Bloch solutions.

\setlength{\textfloatsep}{7pt}

\section{Discussion and conclusion}\label{sec:discussion}
%
Two discrete formulations of the scalar $q(\omega)$ Bloch problem -- in terms of the full field and periodic Bloch factor, respectively -- have been compared. In FF, the Bloch wavenumber enters only through the boundary multiplier $z=\exp(\ii qa)$. In the PF formulation, $q$ appears throughout the wave equation and results in a volume polynomial pencil. At the continuous level, the two descriptions are exactly equivalent and satisfy reciprocal-lattice covariance: $q$ and $q+m\kappa$ correspond to the same full Bloch field. But an important structural difference emerges after the standard discretizations considered here. The FF formulation preserves reciprocal covariance by construction, simply because $z(q + m\kappa) = z(q)$. The corresponding PF discretizations do not, and their eigenproblems can be polluted with inaccurate or nonphysical modes.

The practical recommendation is correspondingly simple: use the FF formulation as the computational default, obtaining the periodic factor by straightforward postprocessing when needed. Schur reduction -- equivalently, a discrete DtN map -- provides an especially efficient implementation of the FF-Bloch problem \cite{Yuan-DtN-BandGaps-2006,TsukermanTrefftzScalarEM2026,SmiarowskiJanaszekTsukermanCEFC2026}.

For the PF formulation, out-of-FBZ roots should not be accepted merely by folding or relabeling them into the FBZ. Reciprocal relabeling leaves the full Bloch field unchanged; physical identification therefore has to precede spectral reporting, and nonphysical roots must be rejected. When an exterior root is a valid reciprocal representative of an FBZ mode, it adds no new physical solution and may simply be omitted from an FBZ presentation.

As a side note related to computational mathematics, this paper can be associated with the broader theme of structure-preserving numerical algorithms. Examples in other areas include (i) symplectic integrators, which preserve the Hamiltonian structure of dynamical systems and thereby improve qualitative long-time fidelity \cite{HairerLubichWanner2006,MarsdenWest2001}; (ii) finite element families and mimetic schemes preserving discrete de~Rham complexes and commuting maps in electrodynamics \cite{Hiptmair2002,ArnoldFalkWinther2006,TeixeiraChew1999,Tonti2001,HymanShashkov1999}.

Apart from the theoretical considerations, the findings and recommendations of this paper are of practical importance because Bloch-periodic factors are standard across several application areas. In photonics they appear in conventional band calculations \cite{Sakoda2005,JohnsonJoannopoulos2001}, topological photonics \cite{LuJoannopoulosSoljacic2014,OzawaEtAl2019,Tsukerman2025}, and non-Hermitian formulations for dispersive and lossy structures \cite{DavancoUrzhumovShvets2007,EngstromRichter2009,NotarosPopovic2015}. The same Bloch-periodic representation is standard in phononic-crystal band calculations \cite{SigalasEconomou1993}. Beyond electrodynamics, cell-periodic Bloch functions underlie $k \cdot p$ and effective-mass theory \cite{LuttingerKohn1955,Kane1957} and topological-insulator band theory \cite{FuKane2007,Vanderbilt2018}.

\bibliographystyle{plainnat}
\bibliography{spurious_Bloch_modes}
\end{document}